\documentstyle[fleqn,twoside]{article}
\def\be{\begin{equation}}
\def\ee{\end{equation}}
\def\bi{\bibitem}

\begin{document}
\title{Is N$\ddot{o}$ther Symmetric Approach Consistent With Dynamical 
Equation In Non-Minimal Scalar-Tensor Theories?} 
\author{A.K.Sanyal$^{*,1}$, 
B.Modak$^{**,2}$} 

\maketitle

\noindent

\begin{center}
${1}$ Dept. of Physics, Jangipur College, Murshidabad,
\noindent
West Bengal, India - 742213. \\
\noindent
${2}$ Dept. of Physics, University of Kalyani,\\
\noindent
India - 741235.
\end{center}

\footnote{\noindent
Relativity and Cosmology Research Centre\\ 
\noindent
Department of Physics, Jadavpur University \\
\noindent
Calcutta - 700 032, India\\
\noindent
Electronic address: * aks@juphys.ernet.in,\\
\noindent
** bmodak@klyuniv.ernet.in}

\noindent
\begin{center}
\bf{\Large{Abstract}}
\end{center}

	The form of the coupling of the scalar field with gravity and the 
potential have been found by applying N$\ddot{o}$ther theorem to 
2-dimensional minisuperspaces in induced gravity model. It has been observed 
that though the forms thus obtained are consistent with all the equations 
$\pounds_{X}L=0$, yet they do not satisfy the field equations for$k = \pm{1}$
, in Robertson-Walker model. It has been pointed out that this is not due to 
the degeneracy of the Lagrangian, since this problem does not appear in 
$k = 0$ case. It has also been shown that though N$\ddot{o}$ther theorem  
fails to extract any symmetry from the Lagrangian of such model for 
$k = \pm 1$, symmetry exists, which can easily be found by studying the 
continuity equation.

\section{\bf{ Introduction}}

	The idea of applying N$\ddot{o}$ther theorem to cosmological 
minisuperspaces is not far too old. In 1990, De Rities etal \cite{r:pr}, 
perhaps for the first time applied it to the homogeneous isotropic 
cosmological models of general theory of relativity with a minimally coupled 
scalar field. Since then, lot of works came out in the literature \cite{r:pl}.
During early nineties, theory of gravitation with nonminimally coupled scalar 
field was supposed to encounter a number of longstanding puzzles of cosmology,
e.g., it was thought to produce successful phase transition for a large class 
of models and help to solve the graceful exit problem on one hand and density 
perturbation on the other \cite{d:pr}. 	
\par

        However, the problem with such model is that, the exact form of the 
coupling of the scalar field with gravity is not known a priori. Capozziello  
etal \cite{c:pl} following \cite{r:pr} tried to find the form of the coupling 
$f(\phi)$ along with the potential $V(\phi)$ for a scalar field $\phi$ by 
applying N$\ddot{o}$ther theorem in induced gravity theory. N$\ddot{o}$ther 
theorem states that, if there exists a vector field X, for which the Lie 
derivative of a given Lagrangian $L$, vanishes $(\pounds_{X}L = 0)$, 
the Lagrangian admits N$\ddot{o}$ther symmetry yielding  a conserved 
current. The idea of the theorem under the present context is to find the 
form of $f(\phi)$ and $V(\phi)$ for which there would exist a vector 
field $X$ that would make $\pounds_{X}L = 0$ and hence would yield a conserved 
current. It is not expected in general that, in all such models there would 
exist certain $f(\phi)$ and $V(\phi)$ and hence $X$, that would make 
$\pounds_{X}L = 0$. Rather, while solving the set of equations 
$\pounds_{X}L = 0$,one might some time encounter inconsistencies which would 
imply that the Lagrangian does not admit N$\ddot{o}$ther symmetry in such 
models. However, it is generally believed that if certain $f(\phi)$ and $V(\phi)$ and hence $X$ 
exists, that satisfy all the equations $\pounds_{X}L = 0$, then the 
Lagrangian admits N$\ddot{o}$ther symmetry.	
\par

         The last statement is not  found to be true in general. In the 
following section we shall review the work of Capozziello etal \cite{c:pl} 
and show, what they might have overlooked, that for induced theory of gravity 
in the Robertson-walker minisuperspace model, there apparently exists, 
N$\ddot{o}$ther symmetry in the sense that all the equations $\pounds_{X}L = 0
$ are satisfied, together with the existence of a conserved current. However, 
the functions $f(\phi)$ and $V(\phi)$ thus obtained are not found to satisfy 
the field equations for $k = \pm 1$. This result perhaps is unique and not been 
encountered earlier. It is true, as pointed out by Capozziello etal 
\cite{c:pl}, that the form of $f(\phi)$  makes the Lagrangian degenerate. 
However, it is known that a degenerate Lagrangian gives rise to a constrained 
system for which Legendre transformation does not exist and Hamiltonian 
dynamics can not be produced unless the constraint is tackled properly 
\cite{k:s}. Further, as far as Lagrangian dynamics is concerned, degeneracy 
introduces underdeterminancy, i.e., the number of field equations are 
less than the number of true degrees of freedom. So degenenary is not 
responsible for the inconsistency mentioned above. 	
\par

We shall further show in this section that for $k = \pm 1$, apparently 
N$\ddot{o}$ther symmetry is found to exist with zero potential but real 
solution exists only for $k = -1$. Due to the degeneracy of the 
Lagrangian, the field equations reduce to one with two field variables in the 
case of $k = -1$. This equation is found to admit inflationary solution.	
\par

At the end of section 2, we shall further show that the same form of 
$f(\phi)$ with a different $V(\phi)$ satisfy the field equations for $k = 0$,
for which all the field equations reduce to one only, viz., the constant 
of motion with two degrees of freedom $a(t)$ and $\phi(t)$. We shall also show
that there exists nontrivial solution of $a(t)$, if certain form of 
$\phi(t)$ is assumed. Hence we conclude that the trivial solution $a(t) = 0,
\dot{\phi(t)} = \phi(t) = 0$, obtained by Capozziello etal \cite{c:pl} is 
not due to the presence of degeneracy rather due to the use of a typical from 
of $f(\phi)$ and $V(\phi)$, which is the out come of N$\ddot{o}$ther 
symmetry, but that do not satisfy the field equations.	
\par

The form of $f(\phi)$ we are talking of so far is only a particular solution 
of a first order non-linear differential equation of $f$ that can not be 
solved in general,to obtain $f$ in closed form. Since we are only interested 
in those Lagrangian in which gravitational coupling is in closed from, so at 
this stage we can always make the comment that, N$\ddot{o}$ther symmetry does 
not exist for such models with $k = \pm 1$, for potentials other than zero. For 
zero potential the symmetry is found to exist only for $k = -1$, which is 
consistent with the dynamical equations. 	
\par

              In section 3 we have chosen a particular form of $f(\phi)$, 
viz., $f(\phi) = \epsilon\phi^2$ keeping a parameter $\epsilon$ arbitrary and 
have shown that the study of the field equation viz., the continuity equation,
reveals the existence of certain symmetry in Lagrangian for which there is a 
constant of motion and which can not be found by studying N$\ddot{o}$ther 
theorem. We had a belief that all dynamical symmetries can be encountered by 
applying N$\ddot{o}$ther theorem. However following the above result such 
belief is now questionable. Section 4 concludes our present work. 

\section{\bf{N$\ddot{o}$ther Symmetry In Robertson-Walker Model}}	

We start with the following action,		
\be
A =\int~~d^4 x (\sqrt{-g} [f(\phi)R- \frac{1}{2} 
(\phi,_{\mu}\phi^{,\mu}-V(\phi))]					 
\ee
which for Robertson-Walker metric, reduces to			       	      	
\be
A = \int[- fa\dot{a}^2- f'a^2\dot{a}\dot{\phi} 
+kaf+\frac{1}{12}a^3 \dot{\phi}^2-\frac{1}{6}a^3 V] dt + surface-term	
\ee
Field equations along with the Hamiltonian constraint equations are    
\be
2\frac{\ddot{a}}{a}+\frac{\dot{a}^2}{a^2}+2\frac{f'a}{fa}\dot{\phi}+\frac{1}{f}[f''+\frac{1}{4}]\dot{\phi}^2+\frac{f'}{f}\ddot{\phi}+\frac{k}{a^2}-\frac{V}{2f}
\ee
\be
\frac{\ddot{a}}{a}+\frac{\dot{a}^2}{a^2}-\frac{\dot{a}\dot{\phi}}{2f'a}-\frac{\ddot{\phi}}{6f'}+\frac{k}{a^2}-\frac{V'}{6f'}=0
\ee                                          
\be
\frac{\dot{a}^2}{a^2}+\frac{f'\dot{a}\dot{\phi}}{fa}-\frac{\dot{\phi}^2}{12f}+\frac{k}{a^2}-\frac{V}{6f}=0
\ee
In the above, $k = 0, \pm 1$ and dot represents derivative with respect to 
proper time, while dash, derivative with respect to $\phi$. In the 
Lagrangian under consideration the configuration space is $Q= (a,\phi)$, 
whose tangent space $TQ= (a,\phi,\dot{a},\dot{\phi})$. Hence the 
infinitestimal 
generator of the N$\ddot{o}$ther symmetry 
is                                            \be       
X=\alpha\frac{\partial}{\partial{a}}+\beta\frac{\partial}{\partial{\phi}}+\dot{\alpha}\frac{\partial}{\partial{\dot{a}}}+\dot{\beta}\frac{\partial}{\partial{\dot{\phi}}}
\ee
The existence of N$\ddot{o}$ther symmetry implies the existence of the 
vector 
field $X$ such that the Lie derivative of the Lagrangian with respect to the 
vector field vanishes i.e.			
\be
\pounds_{X}L = 0								
\ee
This gives an expression of second degree in $a$ and $\phi$, whose 
co-efficient are functions of $a$ and $\phi$ only. Thus to satisfy equation 
(7), we have                                       
\be
\alpha f+a\beta f'+2af\frac{\partial{\alpha}}{\partial{a}}+a^2 
f'\frac{\partial{\beta}}{\partial{a}}=0,
\ee
\be
\alpha-4\alpha'f'+\frac{2}{3}a\beta'=0,
\ee
\be
a\beta f''+[a\frac{\partial{\alpha}}{\partial{a}}+2\alpha+a\beta'] 
f'-\frac{a^2}{6}\frac{\partial{\beta}}{\partial{a}}+2\alpha' f=0,
\ee
\be
k(\alpha f+a\beta f')-\frac{a^2}{2}[\alpha V+\frac{a\beta V'}{3}]=0 .
\ee
The above set of equations can be solved by the method of separation of 
variables. Essentially for $k = \pm 1$, one ends up with a non linear 
differential equation of $f$ which is		
\be
3f'^2 + f = Q f^3
\ee									
where $Q$ is a constant of integration. For $Q = 0$, this equation  can 
immediately be solved to yield		
\be
f = -\frac{1}{12} (\phi-\phi_{0})^2						
\ee
$\alpha,\beta$ and $V$ can also be obtained as a bye product viz.,    
\be
\alpha=\frac{2n}{a(\phi-\phi_{0})^2};\beta=\frac{n}{a^2(\phi-\phi_{0})};V=\lambda(\phi-\phi_{0})^6
\ee
Since, solutions (13) and (14) satisfy the set of equations (8) - (11) 
i.e. equation (7), so one can now claim that the  Lagrangian (2) admits 
N$\ddot{o}$ther  symmetry for the above form of $f$ and $V$. Now for the 
Cartan one form                    
\be
\theta_{L}=\frac{\partial{L}}{\partial{a}}da+\frac{\partial{L}}{\partial{\phi}}d\phi,
\ee
the constant of motion $i_{X}\theta_{L}$ is obtained as, 		    
\be
F=\frac{\frac{d}{dt}[a(\phi-\phi_{0})]}{\phi-\phi_{0}}.
\ee
So far, we have produced nothing new. However, at this stage Capozziello 
etal \cite{c:pl} went  on to find cyclic co-ordinate and have produced a 
trivial solution like $a(t) = 0 = \phi(t)$. It is true that the form of $f$ 
given by (13), makes the Lagrangian (2) degenerate, since its Hessian 
determinant vanishes. Degenerate Lagrangian leads to constrained dynamics for 
which Legendre transformation does not exist and hence Hamiltonian can not be 
expressed. In the context of general relativity degenerate Lagrangian appears 
quite often leading to underdetermined situation, i.e. degrees of freedom 
becomes more in number than the  field equations and relations like equation 
of state etc. are assumed to tackle the situation. However, in the following 
we shall analyse the constraint to inspect the outcome. For this purpose we 
find $f', f''$ and $V'$ from equations (13) and (14) and substitute all 
these together with $f$ and $V$ in equations (3), (4) and (5) to obtain.      
\be
2\frac{\ddot{a}}{a}+\frac{\dot{a}^2}{a^2}+4\frac{\dot{a}\dot{\phi}}{a(\phi-\phi_{0})}-\frac{\dot{\phi}^2}{(\phi-\phi_{0})^2}+2\frac{\ddot{\phi}}{\phi-\phi_{0}}+\frac{k}{a^2}+6\lambda(\phi-\phi_{0})^4 = 0,
\ee
\be
\frac{\ddot{a}}{a}+\frac{\dot{a}^2}{a^2}+3\frac{\dot{a}\dot{\phi}}{a(\phi-\phi_{0})}+\frac{\ddot{\phi}}{\phi-\phi_{0}}+\frac{k}{a^2}+6\lambda(\phi-\phi_{0})^4 = 0,
\ee
\be
\frac{\dot{a}^2}{a^2}+2\frac{\dot{a}\dot{\phi}}{a(\phi-\phi_{0})}+\frac{\dot{\phi}^2}{(\phi-\phi_{0})^2}+\frac{k}{a^2}+2\lambda(\phi-\phi_{0})^4 = 0.
\ee
Now, taking the time derivative of equation (16) and using it to eliminate 
$\ddot{a}$ and $\ddot{\phi}$ terms from equation (17) or (18) one ends up 
with the following equation in either case, viz.,  
\be
\frac{\dot{a}^2}{a^2}+2\frac{\dot{a}\dot{\phi}}{a(\phi-\phi_{0})}+\frac{\dot{\phi}^2}{(\phi-\phi_{0})^2}+\frac{k}{a^2}+6\lambda(\phi-\phi_{0})^4 = 0.
\ee
It is now evident that equations (19) and (20) lead to inconsistency unless 
$\lambda = 0$. This inconsistency can also  be viewed other way round. 
Eliminating $\ddot{a}$ and $k/a^{2}$ terms from equations (3), (4) and (5), 
which is achieved by multiplying equation (4) by 2, subtracting it from 
equation (3) and finally adding it up with equation (5), one obtains 
\be
(3f'^2 +f)[\ddot{\phi}+3\frac{\dot{a}\dot{\phi}}{a}]+(3f'^2 
+f)'\frac{\dot{\phi}^2}{2}+(fV'-2Vf') = 0.
\ee
This is an important relation that we shall come across in the following 
section to find symmetries other than N$\ddot{o}$ther symmetry. However, 
presently we 
observe that for $3f'^2 + f = 0$, which is the equation (12) for $Q = 0$ and 
in particular for solution (13) of $f$, the potential is in the form	
\be
V=\lambda(\phi-\phi_{0})^4 ,						
\ee
$\lambda$ being a constant of integration. It is now apparent that  
for the solution (13) of $f$, field equations are satisfied for a 
potential given by the form (22) and not by the form (14).It can further be 
shown that the potential given by the form (22) does not yield conserved 
current, which implies N$\ddot{o}$ther symmetry does not exist for this 
case. To 
show this, we construct yet another equation by subtracting equation 
(4) from equation (3) viz.,    
\be
\frac{\ddot{a}}{a}+[2\frac{f'}{f}+\frac{1}{2f'}]\frac{\dot{a}}{a}\dot{\phi}+[\frac{f'}{f}+\frac{1}{4f}]\dot{\phi}^2+[\frac{f'}{f}+\frac{1}{6f'}]\ddot{\phi}+\frac{V'}{6f'}-\frac{V}{2f} = 0.
\ee
For the form of $f$ and $V$ given by (13) and (14), one obtains the conserved 
current given by (16). However for any form of the potential other than 
zero, no such conserved current can be obtained, confirming our claim stated 
above. At this point we like to mention that we have come across a very 
important result that has  never been encountered earlier and that is, 
even if a Lagrangian can be found to admit N$\ddot{o}$ther symmetry, it 
might not satisfy the field equations. Truly, in view of equations (19) and 
(20) we can say that it does not satisfy the Hamiltonian Constraint equation in 
particular, which is a typical feature of gravitation, and hence the result 
remained unnoticed. We further like to mention that the trivial solutions 
obtained by Capozziello etal \cite{c:pl} is not due to the degeneracy of the 
Lagrangian, rather due to a wrong choice of potential given by equation (14) 
that does not satisfy the field equations. 
\par
We have already noticed that the conserved current (16) and hence 
N$\ddot{o}$ther symmetry exists for $V=0$, which further satisfies the field 
equations. So let us study the case. For $V = 0$, the derivative of equation 
(16) along with the Hamiltonian  Constraint equation (19) trivially satisfy 
equations (17) and (18). Hence we are left with equation (16) and equation 
(19), with $\lambda = 0$. Now, squaring equation (16)and substituting it in 
equation (19) with $\lambda = 0$, one gets		
\be
F^2 + k = 0								
\ee	
i.e. F is determined by the curvature parameter, which has to be $k = -1$ 
for real solution. Hence, we have come across yet another contradiction that 
though N$\ddot{o}$ther symmetry apparently has been found to exist for $k 
= \pm 1$, but for real solutions only $k = -1$ is allowed. However, even 
for $k = -1$, we have got only one equation viz. (16) to solve for a pair of 
field variables $a$ and $\phi$. This is the out come of a degenerate Lagrangian 
as mentioned earlier. In any case, equation (16) can be solved in 
principle by assuming a form of a or $\phi$. To see whether equation (16) 
admits inflation, let us choose $a = A e^{mt}$, A and m being constants. The 
solution for $\phi$ thus turns out to be   	                 
\be
\phi=C exp[-mt-\frac{F}{Am}exp(-mt)],
\ee
where C is the constant of integration and $\phi_{0}$ has been assumed to be 
zero. As $t \rightarrow  0, \phi \rightarrow c e^{-F/Am}$ and hence becomes 
constant while as $t \rightarrow \infty, \phi \rightarrow  0$. Thus the 
solution of $\phi$ is found to be well behaved. 
\par
We have already shown that for the form of $f$ and $V$ given by relations 
(13) and (22) N$\ddot{o}$ther symmetry does not exist, hence this case can be 
studied to show once again that degeneracy produces underdetermined equations 
and nothing else. In view of relations (13) and (22), field equations (3), (4) 
and (5) reduces to                 
\be
2\frac{\ddot{a}}{a}+\frac{\dot{a}^2}{a^2}+\frac{\dot{a}\dot{\phi}}{a\phi}-\frac{\dot{\phi}^2}{\phi^2}+2\frac{\ddot{\phi}}{\phi}+\frac{k}{a^2}+6\lambda\phi^2=0.
\ee
\be
\frac{\ddot{a}}{a}+\frac{\dot{a}^2}{a}+3\frac{\dot{a}\dot{\phi}}{a\phi}-\frac{\ddot{\phi}}{\phi}+\frac{k}{a^2}+4\lambda\phi^2=0.
\ee
\be
\frac{\dot{a}^2}{a^2}+2\frac{\dot{a}\dot{\phi}}{a\phi}+\frac{\dot{\phi}^2}{\phi^2}+\frac{k}{a^2}+2\lambda\phi^2=0.
\ee
where, for the sake of simplicity we have chosen $\phi_{0} = 0$. We further 
restrict ourself to the case $k = 0$. Now multiplying equation (28) by 3 
and subtracting it from equation (26) one obtains,  
\be
\frac{\ddot{a}}{a}-\frac{\dot{a}^2}{a^2}+\frac{\ddot{\phi}}{\phi}-2\frac{\dot{\phi}^2}{\phi^2}-\frac{\dot{a}\dot{\phi}}{a\phi}=0,
\ee
which further yields  
\be
\frac{\dot{a}}{a}+\frac{\dot{\phi}}{\phi}=m\phi,
\ee
where, m is the constant of integration. Using equation (26) in 
equations(22), (23) and (24) one obtains only a single outcome, which is a 
restriction of m, viz.,		
\be
m^2 + 2\lambda  = 0.									
\ee
So m is determined. Equation (31) implies that for real solution $\lambda$ 
should be negative. Hence we end up with a single equation viz., (30). 
This is what we have claimed earlier, that degeneracy in Lagrangian produces
an underdetermined situation, viz., one equation for two degrees of 
freedom, in the present context. 
\par
One can now try to solve equation (30) for the  scale factor by choosing a 
solution for $\phi$ with the requirement that the early universe was 
dominated by the scalar field while it vanishes asymptotically. For 
such requirement one can choose $\phi$ in the following form
\be
\phi=e^{-p^2 {t}},
\ee
where $p^2$ is a constant. Under such assumption given by equation (32) the 
scale factor takes the following form	               
\be
a=le^{[p^2{t}-\frac{m}{p^2}e^{-p^2{t}}]},
\ee
where $l$ is yet another constant of integration. The above solution shows 
that the scale factor vanishes as $t \rightarrow -\infty$, while it has 
got a constant value at $t = 0$ and exponentially increases with t. 
	
\section{\bf{No Other Symmetry In Robertson-Walker Model?}}	

In the last section we have observed that for $k = \pm 1$, the 
N$\ddot{o}$ther symmetry that
was supposed to exist, actually does not, since the required forms of $f$ 
and $V$ for the existence of N$\ddot{o}$ther symmetry do not satisfy the 
field 
equations except for a very special case viz., $V = 0$ and $k = - 1$. Such 
a typical behaviour appears for the first time and perhaps, Hamiltonian 
constraint, 
which is a special feature of gravitation is responsible for such 
contradiction. The question that arises, whether $k = \pm 1$ admits any 
other 
symmetry or not. As far as N$\ddot{o}$ther symmetry is concerned we can say 
that the 
techniques we have followed to find such symmetry will not yield anything 
further. This is because equation (12) cannot be solved to obtain $f$ in  
closed form. However, still there might exist symmetry which can be studied  
from the field equations only. To show this we choose a form of equations 
$f = -\epsilon \phi^2$, keeping $\epsilon$ arbitrary and study whether any 
symmetry exists for 
$\epsilon \ne 1/12$. Thus the action is now
\be
A = \int d^4 x \sqrt{-g} [-\epsilon\phi^2 R- 
\frac{1}{2}(\phi,_{\mu}\phi^{,\mu}-V(\phi))],    				        
\ee
which reduces to               
\be
A = \int[6\epsilon a \dot{a}^2 \phi^2 + 12\epsilon a^2 
\phi\dot{a}\dot{\phi}- 6\epsilon ka\phi^2 + \frac{1}{2}( a^3 \dot{\phi}^2- 
a^3 V(\phi))] dt + surface-term.  
\ee
Thus the field equations are
\be
2\frac{\ddot{a}}{a}+\frac{\dot{a}^2}{a^2}+4\frac{\dot{a}\dot{\phi}}{a\phi}+[2-\frac{1}{4\epsilon}]\frac{\dot{\phi}^2}{\phi^2}+2\frac{\ddot{\phi}}{\phi}+\frac{k}{a^2}+\frac{V}{2\epsilon\phi^2}=0,
\ee
\be
\frac{\ddot{a}}{a}+\frac{\dot{a}^2}{a^2}+\frac{\dot{a}\dot{\phi}}{4\epsilon a\phi}+\frac{\ddot{\phi}}{12\epsilon\phi}+\frac{k}{a^2}+\frac{V'}{12\epsilon\phi}=0,
\ee
\be
\frac{\dot{a}^2}{a^2}+2\frac{\dot{a}\dot{\phi}}{a\phi}+\frac{\dot{\phi}^2}{12\epsilon \phi^2}+\frac{k}{a^2}+\frac{V(\phi)}{6\epsilon \phi^2}=0.
\ee
One can now proceed to find N$\ddot{o}$ther symmetry for the Lagrangian 
(35), but then 
it will end up with the same result, ie., for $k = \pm 1$, N$\ddot{o}$ther 
symmetry 
exists for $\epsilon = 1/12$, which not only makes the Lagrangian degenerate 
but also restricts $V$ in such a way $(V = \lambda \phi^6)$, that field 
equations are not satisfied. However, one can still study the 
possibility of existence of some other form of symmetry, in view of the   
the continuity equation, which is obtained  by 
multiplying  equation (37) by 2 and then subtracting it from equation (31) 
and finally adding it up with equation (38). The  equation thus formed is the 
same equation (21) but for $f = - \epsilon \phi^2$ viz.,        
\be
(12\epsilon-1)[\frac{\ddot{\phi}}{\phi}+\frac{\dot{\phi}^2}{\phi^2}+3\frac{\dot{a}\dot{\phi}}{a\phi}]+\frac{\phi V'-4V}{\phi^2}=0.
\ee
It is clear from the above equation that if one chooses $\epsilon= 1/12$ 
then $\phi V' - 4V = 0$ i.e., $V = \lambda \phi^4$, which is the same old 
result. On 
the contrary if one starts with a quartic potential i.e. $V = \lambda 
 \phi^4$ and restricts oneself to such situation that the Lagrangian does 
not turn out to be degenerate i.e.$\epsilon \ne 1/12$ , then for any 
arbitrary $\epsilon$ one would obtain
\be
\frac{\ddot{\phi}}{\phi}+\frac{\dot{\phi}^2}{\phi^2}+3\frac{\dot{a}\dot{\phi}}{a\phi}=0,
\ee
i.e.,
\be
a^3 \phi\dot{\phi}=constant.
\ee
Hence we obtain a constant of motion, which implies that the Lagrangian has 
got a symmetry for all $k = 0, \pm 1$, that does not make it degenerate. 
Clearly, $f = \epsilon \phi^2$ does not  satisfy equation (12) for 
arbitrary $Q$, i.e., the symmetry obtained just by studying the 
continuity equation does not belong to N$\ddot{o}$ther class. This result 
implies 
that not all (dynamical?) symmetries of Lagrangian are revealed  by studying 
N$\ddot{o}$ther theorem.
\section{\bf{Conclusion}}

The two very important outcomes of this paper are as follows : (1) The  
Lagrangian (2) seems to admit N$\ddot{o}$ther symmetry $\pounds_{x} L = 0$  
for $(k = \pm 1)$     
$f =  - 1/12 (\phi - \phi_{0})^2$ and $V = \lambda (\phi - \phi_{0})^6$. 
However these forms of 
$f$ and $V$ do not satisfy the field equations except for very special case 
viz., 
$V = 0$ and $k = -1$ yielding nontrivial solutions of the field equations. 
For $f = - 1/12 (\phi - \phi_{0})^2$, the field equations are satisfied for 
$V =\lambda (\phi - \phi_{0})^4$, but this does not yield a conserved 
current for $k = \pm 1$. 
The degeneracy of the Lagrangian is not responsible for such contradiction, 
since for $k = 0, f = - 1/12 (\phi - \phi_{0})^2$  and $V = 
\lambda (\phi - \phi_{0})^4$  we have obtained a 
constant of motion,  viz., equation (30) which does not resemble with 
N$\ddot{o}$ther symmetry. The constraint due to the degenerate Lagrangian in 
this case leads 
to a single equation with a pair of field variables $a$ and $\phi$, that 
can be 
solved in principle assuming a particular form of $\phi$ and the solution is 
found 
to be nontrivial. Such situation arises quite often in general theory of 
relativity and the problem is tackled by assuming equation of state or some 
other condition. Hence we conclude that the trivial solutions viz., $a(t) 
= 0$ 
and $\phi(t) = 0$, as obtained by Capozzeillo etal \cite{c:pl} is not 
due to the 
constraint imposed by degenerate Lagrangian as pointed out by them, rather 
due to the choice  of the potential $V = \lambda \phi^6$, that does not 
satisfy the field equations. 
\par
Situation, like N$\ddot{o}$ther symmetry exists but field equations are not 
satisfied 
has not perhaps been encountered before. The reason for such contradiction 
might be due to the presence of Hamiltonian constraint equation, since we 
have observed that this equation in particular is not satisfied by the 
symmetry.
\par
(2) Though the Lagrangian (2) does not admit N$\ddot{o}$ther symmetry 
$\pounds_{x} L = 0$, yet there exists symmetry that can  be studied  just 
from the 
continuity equation. It has been shown in section 3, that for $k = 0, \pm 1; 
V = \lambda \phi^4$ and $f = - \epsilon \phi^2, \epsilon \ne 0, \ne 1/12$ 
being arbitrary, there exists a constant 
of motion and hence a symmetry in the Lagrangian. For $\epsilon \ne 0$ and 
$\ne 1/12$, the Hessian 
determinant does not vanish and hence Lagrangian is not degenerate. This 
symmetry is not of N$\ddot{o}$ther class since such $f$ and $V$ do not 
satisfy equation 
(12). This implies that even though N$\ddot{o}$ther theorem fails to extract 
any symmetry 
from the Lagrangian, continuity equation might help to see whether symmetry 
exists in any other form. Hence we conclude that the N$\ddot{o}$ther 
theorem is not the last word on (dynamical ?) symmetry.

\begin{thebibliography}{20}
\bi{r:pr} R. de Ritis et al. (1990) Phys. Rev D42, 1091 
\bi{r:pl} R. de Ritis et al. (1991) Phys Lett A 161, 230, M. Demianski et 
al, (1991) Phys Rev D44, 3136 
\bi{d:pr} D. La and P. J. Steinhardt, (1989) Phys. Rev. lett 62, 376, D. La, 
P. J. Steinhardt and E. W. Bertschinger, (1989) Phys. Lett B 231, 231, 
A. D. Linde, (1990) Phys. Lett B 238, 160.
\bi{c:pl} S. Capozziellio and R. de  Ritis. (1993) Phys. Lett A 177, 1, S. 
Capozziello, R. de Ritis and P. Scudellaro (1994) Phys Lett A 188, 130,  
S. Capozziello et al (1996) Nuovo cimento B 109, 159, S. Capozziellio and 
R. de  Ritis, (1994) Class. Quantum Grav 11, 107. 
\bi{k:s} K. Sundermeyer, (1982) Constrained Dynamics, Springer-Verlag.

\end {thebibliography}

\end{document}